%% file: main.tex
\documentclass[runningheads]{llncs}
\usepackage{graphicx}
\usepackage{orcidlink}
\usepackage{amsfonts}
\usepackage{mathtools}
\usepackage{subcaption}
\usepackage{balance}

\begin{document}

\title{Efficient 3D Brain Tumor Segmentation with Axial-Coronal-Sagittal Embedding}
\titlerunning{Efficient 3D BraTS with ACS Embedding}

\author{Tuan-Luc Huynh\inst{1,2}\orcidlink{0000-0002-2481-0724} 
\and Thanh-Danh Le\inst{1,2}\orcidlink{0000-0001-9181-5800}
\and \\ Tam V. Nguyen\inst{3}\orcidlink{0000-0003-0236-7992} 
\and Trung-Nghia Le\inst{1,2}\orcidlink{0000-0002-7363-2610}
\and Minh-Triet Tran\inst{1,2}\orcidlink{0000-0003-3046-3041}}

\institute{
    University of Science, VNU-HCM, Ho Chi Minh City, Vietnam \and
    Vietnam National University, Ho Chi Minh City, Vietnam \and
    University of Dayton, United States
}
\authorrunning{Huynh et al.}

\maketitle              % typeset the header of the contribution

\vspace{-4mm}
\begin{abstract}
In this paper, we address the crucial task of brain tumor segmentation in medical imaging and propose innovative approaches to enhance its performance. The current state-of-the-art nnU-Net has shown promising results but suffers from extensive training requirements and underutilization of pre-trained weights. To overcome these limitations, we integrate Axial-Coronal-Sagittal convolutions and pre-trained weights from ImageNet into the nnU-Net framework, resulting in reduced training epochs, reduced trainable parameters, and improved efficiency.
Two strategies for transferring 2D pre-trained weights to the 3D domain are presented, ensuring the preservation of learned relationships and feature representations critical for effective information propagation. Furthermore, we explore a joint classification and segmentation model that leverages pre-trained encoders from a brain glioma grade classification proxy task, leading to enhanced segmentation performance, especially for challenging tumor labels. Experimental results demonstrate that our proposed methods in the fast training settings achieve comparable or even outperform the ensemble of cross-validation models, a common practice in the brain tumor segmentation literature.

\end{abstract}

\input{mainchaps/introduction}
\input{mainchaps/related_work}
\input{mainchaps/method}
\input{mainchaps/experiments}
\input{mainchaps/conclusion}

\bibliographystyle{splncs04}

 \vspace{-2mm}
\bibliography{references}
\end{document}

%% file: mainchaps/introduction.tex
\vspace{-8mm}
\section{Introduction}
\label{sec:introduction}
\vspace{-2mm}
Brain tumor segmentation in medical imaging is a critical task with significant applications in diagnosis and treatment planning. nnU-Net~\cite{isensee_nnunet_miccai_2021}, the winning solution in BraTS2020~\cite{menze_tmi_2015,bakas2017advancing,bakas_arxiv_2019}, has shown promising performance by incorporating heuristic rules for pre-processing and network architectures. However, it encounters limitations, including the need for extensive training and the lack of utilization of pre-trained weights. Leveraging pre-trained weights from large-scale image datasets, such as ImageNet~\cite{deng_imagenet_cvpr}, has been proven to be valuable in terms of achieving competitive results in various computer vision tasks; Therefore, using these pre-trained weights on downstream tasks to leverage learned representations as a good initialization is desirable instead of random initialization. To address these limitations in brain tumor segmentation, we propose novel approaches based on transfer learning with pre-trained weights and proxy task training to enhance both efficiency and performance of nnU-Net~\cite{isensee_nnunet_nature_2021,isensee_nnunet_miccai_2021}.

Our inspiration for this work stems from a presentation at the Brainlesion workshop during MICCAI2023, where the successful use of Axial-Coronal-Sagittal (ACS) convolutions~\cite{yang_acs_jbhi_2021} in the GanDLF framework~\cite{pati_gandlf_nature_2023} led to superior performance on the BraTS2020 dataset with fewer epochs, in contrast to claims by Isensee~\emph{et al.}~\cite{nnunet_github}. Additionally, upon careful examination, we observe striking similarities between the nnU-Net generated network and ResNet18~\cite{he_cvpr_2016}, providing an opportunity to leverage the valuable pre-trained weights of ResNet18 while preserving the integrity of the nnU-Net framework. To this end, we propose an innovative integration approach incorporating ACS convolutions into the nnU-Net framework~\cite{isensee_nnunet_nature_2021}. We introduce two distinct methods for transferring pre-trained 2D weights to the 3D domain, focusing on preserving nnU-Net's network topology during initialization while retaining critical learned relationships and feature representations essential for effective information propagation.

Furthermore, we explore a joint classification and segmentation model for enhanced performance. This approach leverages a pre-trained encoder on a brain gliomas grade classification proxy task, guiding the segmentation task and leading to improved overall performance. Through these comprehensive contributions, we aim to push the boundaries of brain tumor segmentation, opening up new possibilities for efficient and accurate medical image analysis. Our contributions are summarized as follows:

\vspace{-0mm}
\begin{itemize}
\item Explore nnU-Net's efficacy in fast training settings, reducing required training epochs for brain tumor segmentation.
\item Introduce Axial-Coronal-Sagittal (ACS) convolutions and propose two integration strategies to seamlessly transfer 2D pre-trained weights (e.g., ImageNet) to the 3D domain within nnU-Net's network topology, enhancing training efficiency, and preserving crucial learned relationships and feature representations for effective information propagation.
\item Propose a joint classification and segmentation nnU-Net model leveraging pre-trained proxy tasks for distinguishing high-grade and low-grade gliomas, enhancing segmentation for the challenging enhancing tumor labels.
\item Achieve competitive or better performance than baseline models while maintaining efficiency through reduced training epochs and trainable parameters.
\end{itemize}

%% file: mainchaps/related_work.tex
\section{Related work}
Since its inception, U-Net~\cite{ronneberger_unet_miccai_2015} has become the \textit{de facto} standard for medical image segmentation with its iconic encoder and decoder architecture. Recent BraTS challenge solutions~\cite{menze_tmi_2015,bakas_arxiv_2019,baid_arxiv_2021} have prominently relied on CNNs using the U-shaped encoder-decoder design~\cite{myronenko_miccai_2019,jiang_cascaded_unet_miccai_2020,isensee_nnunet_miccai_2021,luu_miccai_2022,zeineldin_miccai_2022}. Notably, Isensee \emph{et al.} secured victory in the BraTS2020 challenge with the nnU-Net framework~\cite{isensee_nnunet_nature_2021}, which features an automated configuration mechanism. The winning solution of the subsequent year~\cite{luu_miccai_2022} further built upon nnU-Net's achievements. Despite transformer-based methods like TransBTS~\cite{wang_transbts_miccai_2021} and SwinUNetR~\cite{swinunetr} showing comparable performance, CNNs continue to dominate in medical image segmentation.

In the existing literature, a common strategy involves laborious cross-validation, assembly, and training of multiple models from scratch. To address the efficiency challenge, transfer learning has garnered attention. One notable attempt in this direction is Med3D~\cite{chen_med3d_arxiv_2019}, which offers 3D pre-trained weights. However, it is crucial to acknowledge that the scale of its pre-trained data remains incomparable to the vast 2D natural image datasets commonly utilized in transfer learning.

Another approach, Model Genesis~\cite{zhu_modelgenesis_mia_2021}, leverages self-supervised methods on 3D medical images. Although innovative, these methods have yet to surpass the performance of widely explored fully-supervised approaches, which span over a decade of research in the field.

Our work posits that the full potential of efficient training and fine-tuning in the brain tumor segmentation problem is yet to be fully realized, particularly concerning large-scale 3D pre-trained models.

%% file: mainchaps/method.tex
%\vspace{-4mm}
\section{Proposed methods}
\label{sec:proposed_methods}

\subsection{ACS nnU-Net}
\label{sec:acs_nnunet}

Our work builds on Sarthak Pati's insights~\cite{pati_gandlf_nature_2023} and aims to explore the benefits of using pre-trained weights in the nnU-Net framework for 3D medical image segmentation. nnU-Net has shown strength in semantic segmentation but lacks specialized pre-trained weights for 3D medical images. To overcome this, we propose integrating 2D pre-trained ACS convolutions into nnU-Net. This integration seeks to demonstrate significant performance improvements in demanding biomedical image segmentation tasks. However, a major challenge is transferring these pre-trained weights while preserving nnU-Net's core design principle of adapting the network topology for each dataset. Maintaining the integrity of the generated network is essential to address this issue. For further details of ACS convolutions, we refer readers to~\cite{yang_acs_jbhi_2021}.

\subsubsection{ACS-ResNet18 nnU-Net}
\label{sec:acs_resnet18_nnunet}
\begin{figure}[t!]
\centering
\includegraphics[width=1\textwidth]{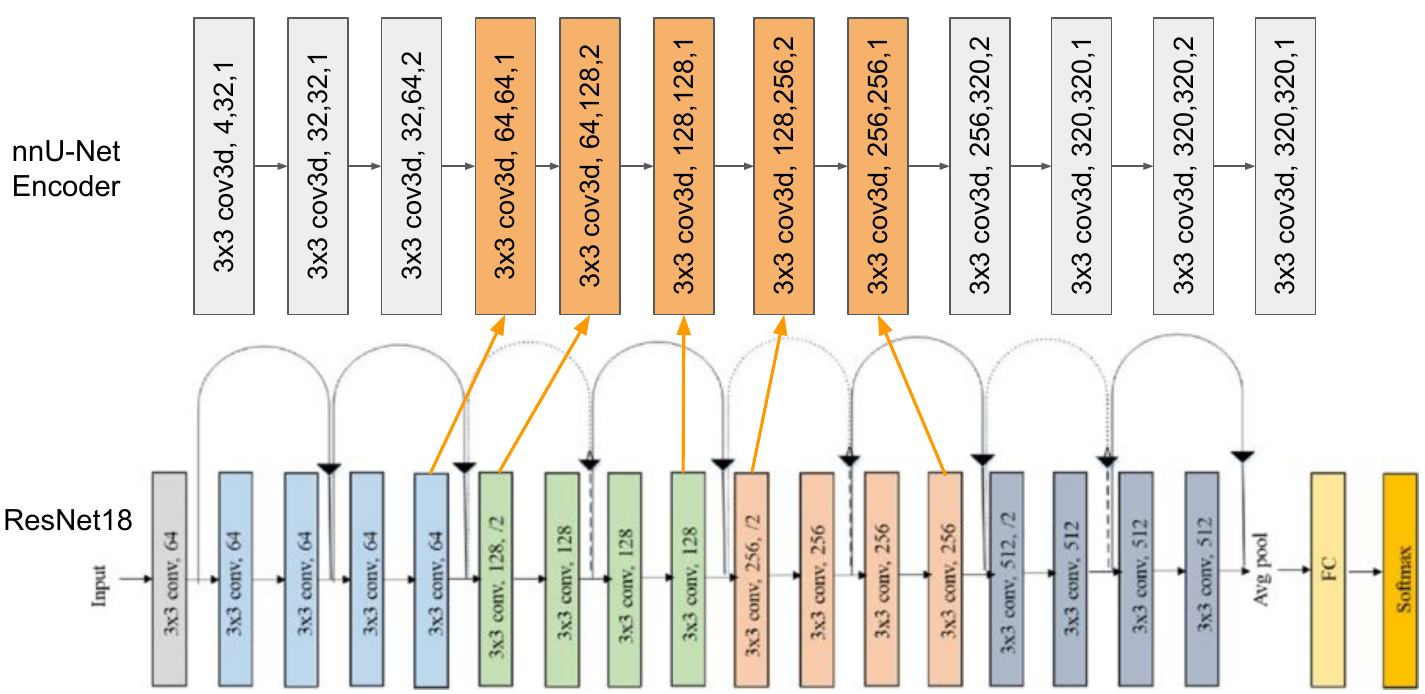}
\caption{Schematic design of our ACS nnU-Net encoder. Convention: ``kernel-size, input channels, output channels, stride'' in each block.}
\vspace{-2mm}
\label{fig:acsresnet18}
\end{figure}

We integrate ACS convolutions into nnU-Net by leveraging similarities with ResNet18~\cite{he_cvpr_2016}. This allows us to use ResNet18 as pre-trained weights for the encoder, as shown in Figure \ref{fig:acsresnet18}. We replace 3D convolutions in the decoder with ACS convolutions, reducing parameters and ensuring symmetry with the encoder, leading to optimized performance.

Our approach uses partial transfer learning, seamlessly integrating 2D domain pre-trained weights with ACS convolutions in nnU-Net. Careful selection and initialization of the corresponding layers preserve critical learned relationships and feature representations, facilitating the effective propagation of information. We prioritize connected pre-trained layers, initializing nnU-Net's specific convolutions with corresponding layers in ResNet18 to enhance integration.

We follow nnU-Net's default Kaiming normal initialization for non-initialized layers ~\cite{he_kaiminginit_iccv_2015}. Our comprehensive approach, encompassing encoder and decoder, exploits pre-trained weights, ACS convolutions, and connected CNN pre-trained layers to boost nnU-Net's performance.

\subsubsection{ACS-All-ResNet18 nnU-Net}
\label{sec:acs_all_resnet18_nnunet}
To increase the availability of pre-trained weights, we introduce a second variant called ACS-All for the nnU-Net. Unlike the previous variant, which only transferred pre-trained weights for convolutional layers with matching properties, ACS-All utilizes sets of connected pre-trained layers in ResNet18 to initialize corresponding layers in the nnU-Net encoder.

\begin{figure}[!b]
\centering
\vspace{-2mm}
\includegraphics[width=1\textwidth]{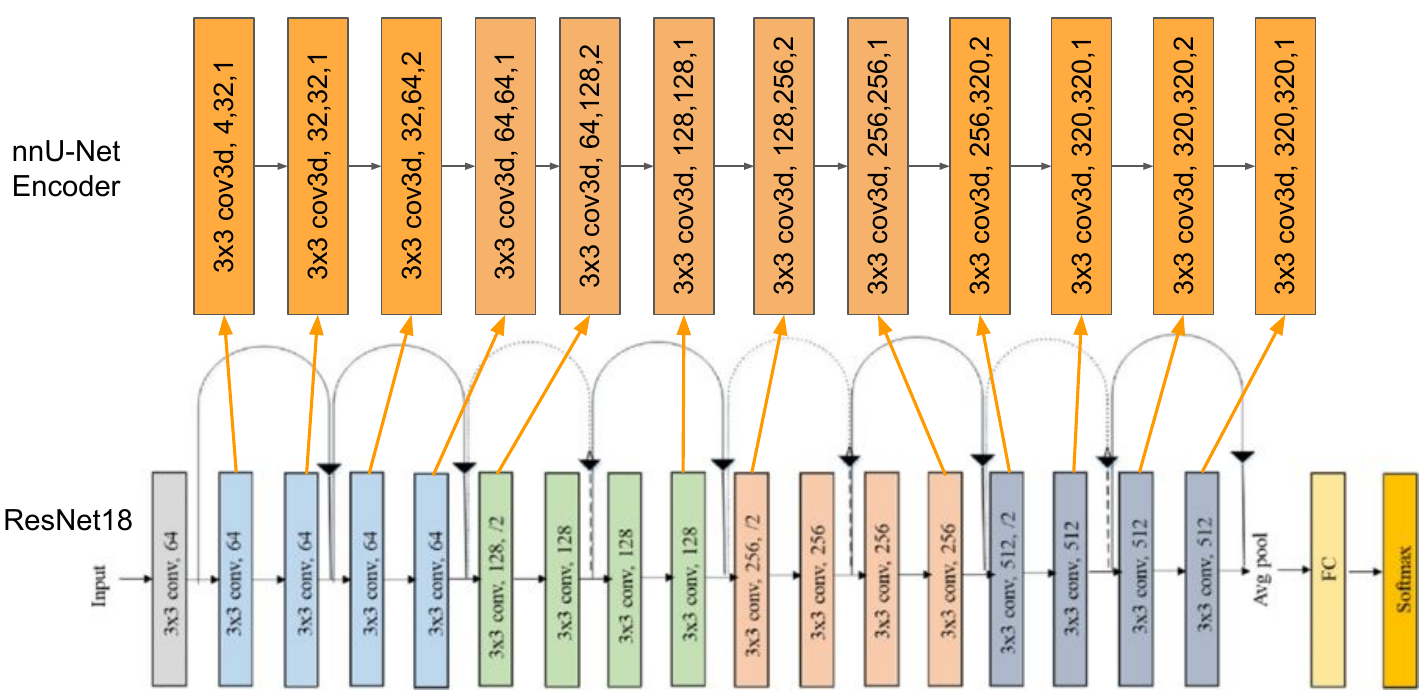}
\caption[Schematic design of our ACS-All nnU-Net encoder]{Schematic design of our ACS-All nnU-Net encoder.}
\label{fig:acsresnet18_all}
\end{figure}

During initialization, ACS-All addresses cases where desired input and output channel sizes differ from those in the pre-trained layers. For example, when the nnU-Net encoder starts with a $4\times32$ layer, which is not directly present in ResNet18, the variant employs a slicing and reshaping strategy. By slicing the first three connected $64\times64$ layers, the variant achieves a seamless initialization process, matching the desired shapes for the nnU-Net encoder layers ($4\times32$, $32\times32$, $32\times64$).
Similarly, for larger input and output channels like $320\times320$, ACS-All benefits from the largest available $512\times512$ pre-trained layer in ResNet18. Selecting the first 320 channels from the pre-trained weights ensures that the most relevant and connected information is used to initialize the nnU-Net layer.
Figure \ref{fig:acsresnet18_all} provides a schematic visualization of the integration process, clearly representing the ACS-All variant initialization procedure.

\subsection{Joint classification and segmentation (JCS) nnU-Net}
\label{sec:jcs}
To preserve performance with limited training epochs, we leverage pre-training the nnU-Net encoder with a glioma grade classification proxy task using HGG/LGG metadata. This equips the encoder with a comprehensive understanding of brain MRI characteristics, enhancing segmentation performance even in cases with minimal enhancing tumor or none. This pre-training approach shares the same motivation as the ACS-based nnU-Net variants, optimizing integration and leveraging pre-trained knowledge for improved performance in challenging biomedical image segmentation tasks like the BraTS dataset~\cite{menze_tmi_2015,bakas2017advancing,bakas_arxiv_2019}.

\begin{figure}[!b]
\centering
\includegraphics[width=0.9\textwidth]{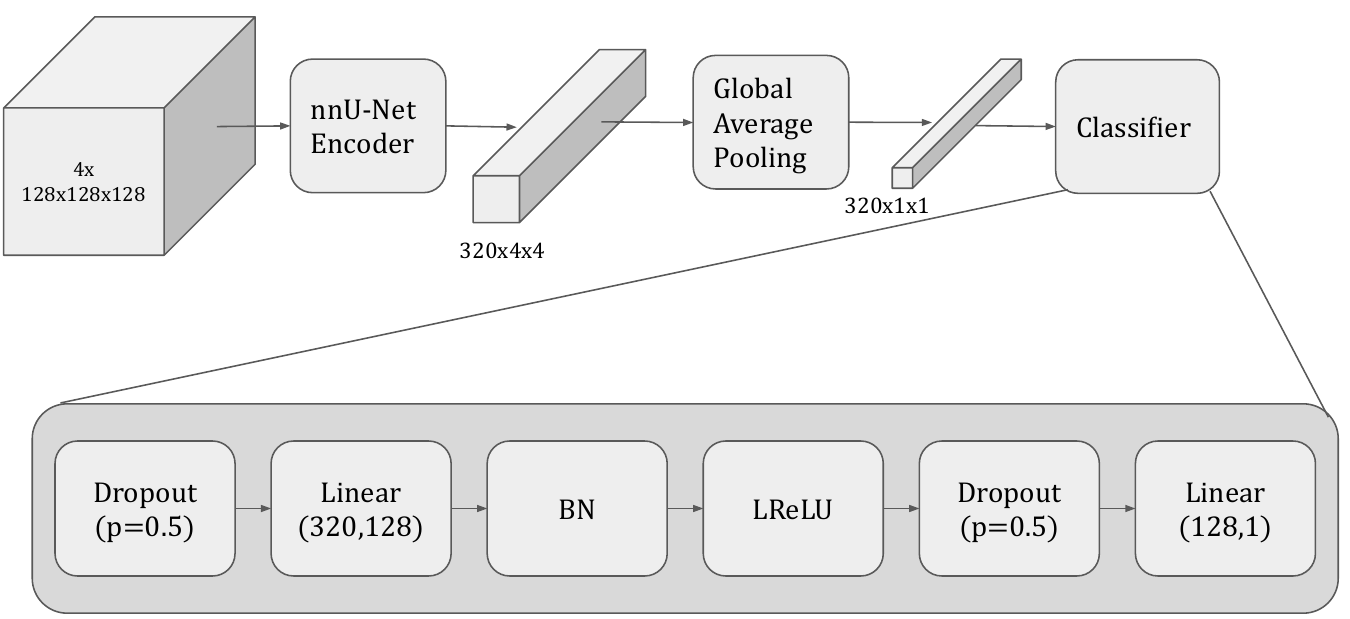}
\caption{HGG-LGG nnU-Net classifier.}
\label{fig:hgg_lgg_classifier}
\end{figure}

To perform the proxy task pre-training, the encoder is detached from the nnU-Net architecture, and a classifier head is added (Figure \ref{fig:hgg_lgg_classifier}). Spatial CNN features are aggregated using global average pooling, followed by fully connected layers with LeakyReLU activation. Dropout layers are incorporated for regularization during training. The learned representations from the proxy task are then leveraged for the tumor segmentation task within the original nnU-Net encoder-decoder architecture.

Directly substituting the nnU-Net's encoder with the HGG-LGG encoder's classification component does not significantly improve performance compared to the baseline due to the inherent task gap between classification and segmentation. Instead, we propose a strategic adaptation by incorporating the HGG-LGG nnU-Net encoder's classification capabilities as a frozen component within the Joint Classification and Segmentation (JCS) model, inspired by Wu~\emph{et al.}~\cite{wu_jcs_tip_2021}. This unified framework combines classification and segmentation strengths, improving tumor segmentation results' accuracy and robustness.

The decision to freeze the classification branch is partly due to limited training resources and overall efficiency. In the JCS framework, we achieve a powerful and balanced approach by combining classification features from the glioma grade classification task to enhance the segmentation task performance.

\begin{figure}[t]
\centering
\includegraphics[width=0.9\textwidth]{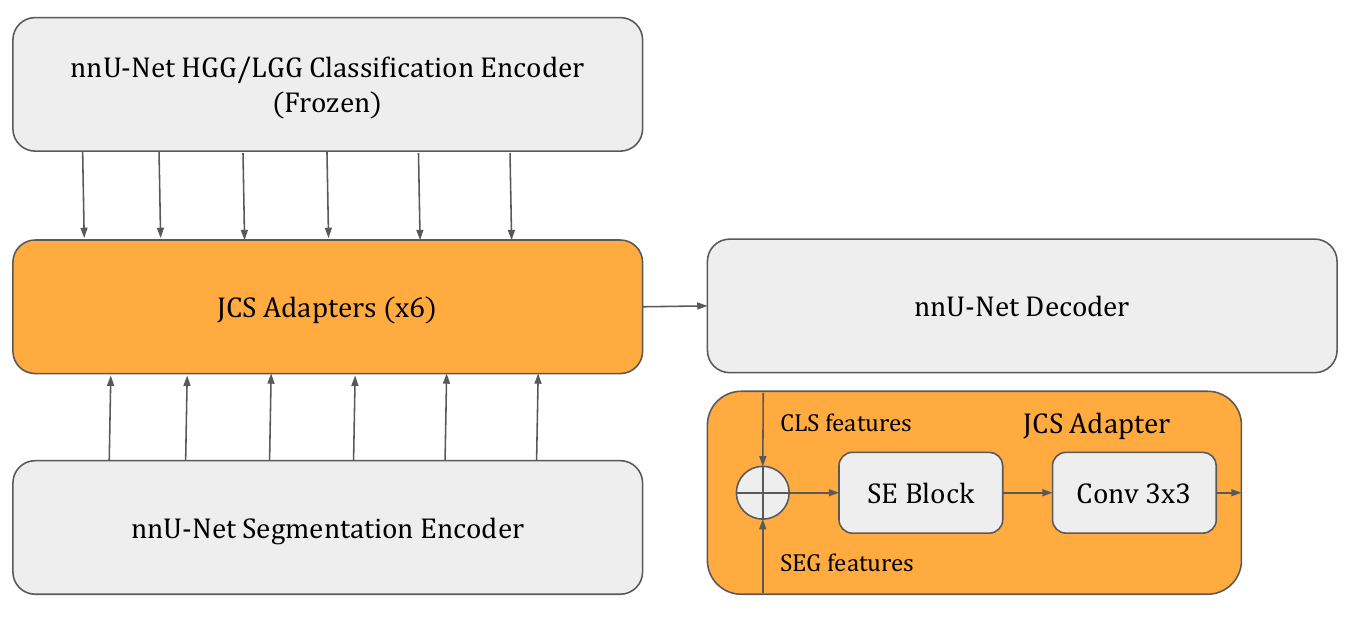}
\caption[JCS nnU-Net]{Schematic view of the JCS nnU-Net.}
\label{fig:jcs_nnunet}
\end{figure}

Figure \ref{fig:jcs_nnunet} illustrates the schematic view of our proposed Joint Classification and Segmentation (JCS) nnU-Net architecture. JCS Adapter, inspired by Wu~\emph{et al.}~\cite{wu_jcs_tip_2021}, is used to seamlessly integrate classification and segmentation while bridging the gap between the two tasks. Given that the nnU-Net encoder consists of six stages, we employed six adapters, each tailored to a specific stage, with variations in input and output channels. The JCS Adapter takes two inputs, classification features (CLS) and segmentation features (SEG), which are concatenated before being passed through a subsequent SE block~\cite{hu_squeezeandexcitation_cvpr2018} and a $3 \times 3$ convolution layer.

%% file: mainchaps/experiments.tex
\section{Experimental Setup}
\subsection{Dataset}
\vspace{-2mm}
We focus on the BraTS2018 dataset~\cite{menze_tmi_2015,bakas2017advancing,bakas_arxiv_2019} due to its relevance in evaluating our research. Additionally, we assess the transferability and robustness of our methods on the more recent BraTS2020 data set~\cite{menze_tmi_2015,bakas2017advancing,bakas_arxiv_2019}. Utilizing both datasets allows us to explore performance in different contexts, advancing brain tumor segmentation in medical image analysis.

\vspace{-4mm}
\subsubsection{BraTS2018}
\label{brats2018}
The BraTS2018 dataset comprises 285 cases, with 66 cases for validation and 191 for testing. It includes multimodal pre-operative MRI scans of High-Grade Gliomas (HGG) and Low-Grade Gliomas (LGG) from 19 institutions. The dataset provides Native (T1), Post-contrast T1-weighted (T1Gd), T2-Weighted (T2), and T2 Fluid Attenuated Inversion Recovery (FLAIR) modalities, with annotations for three sub-regions: Necrotic and non-enhancing tumor (NCR/NET, 1), Peritumoral edema (ED, 2), and GD-enhancing tumors (ET, 4). To maintain resource efficiency without sacrificing representation, we exclusively use the first fold of nnU-Net's 5-fold cross-validation split for our experiments.

\vspace{-4mm}
\subsubsection{BraTS2020}
\label{brats2020}
The BraTS2020 dataset is based on BraTS2018, offering 369 training cases, 125 validation cases, and 166 test cases. It includes a more diverse collection of multimodal MRI scans and refines the naming convention. Similar to BraTS2018, our experiments solely focus on the first fold of nnU-Net's 5-fold cross-validation split for BraTS2020, ensuring consistent and representative evaluations of proposed methods.

\subsection{Evaluation metrics}
\subsubsection{Dice score (Sørensen–Dice coefficient)}, also known as Dice similarity coefficient (DSC), or simply Dice score, quantifies the agreement between segmented regions and actual regions of interest. It is computed by comparing true positives to the sum of pixels in both the predicted and ground-truth masks. The Dice coefficient is defined as follows:

\begin{equation}
    DSC = \frac{2\text{TP}}{2\text{TP} + \text{FP} + \text{FN}}
\end{equation}
\subsubsection{Hausdorff distance 95\% (HD95)} calculates the maximum distance of a set to the nearest point in another set. 95\% Hausdorff distance is based on calculating the 95th percentile of the distances between the boundary points in X and Y. The formula for Hausdorff distance is given by:

\begin{equation}
d_H(X,Y)=\max{d_{XY}, d_{YX}}=\max{\max_{x\in X}\min_{y\in Y}d(x,y), \max_{y\in Y}\min_{x\in X}d(x,y)}.
\end{equation}

\subsection{Implementation details}
\label{sec:implementation}
All segmentation experiments are conducted using the nnU-Net framework \cite{isensee_nnunet_nature_2021} with provided planning files. We use a batch size of 4 and train for 50 epochs on a single Nvidia Quadro RTX 5000 GPU with 16GB VRAM. The JCS models are trained on an Nvidia RTX A5000 GPU with 24GB VRAM. HGG/LGG classifiers undergo 100 epochs of training with a batch size of 5, learning rate of 0.001, stratified sampling strategy, and PyTorch's BCEWithLogitsLoss with a pos\_weight parameter set to the LGG-to-HGG ratio.

\subsection{Implementation variants} 

We represent our models, with region-based training~\cite{isensee_nonewnet_miccai_2019,isensee_nnunet_miccai_2021} by default, using different variants:
\vspace{-2mm}

\begin{itemize}
    \item \textbf{BN}: Batch normalization instead of instance normalization.
    \item \textbf{DA}: BraTS-specific intensive data augmentation~\cite{isensee_nnunet_miccai_2021}.
    \item \textbf{Post}: apply a post-processing strategy using empirical thresholds of 200 voxels for BraTS2018 and 1000 voxels for BraTS2020.
    \item \textbf{CV}: 5-fold cross-validation and default post-processing as nnU-Net's training scheme. The final result is an ensemble of these five models.
\end{itemize}
\vspace{-4mm}

\section{Experimental results}

\subsection{Quantitative results}
\label{sec:quan_results}

\subsubsection{BraTS2018}

\paragraph{Baselines:} 
\vspace{-2mm}
Table \ref{tab:brats2018_baselines} presents the baselines examined in our fast training setup. We exclusively train and evaluate our model on the first fold of the 5-fold cross-validation split file. This approach yields a mean Dice score of 0.85394 and a mean HD95 of 5.70207, establishing it as our baseline evaluation (in bold). Replacing instance normalization with batch normalization marginally reduces the Mean Dice score in the Baseline+BN variant, but significantly improves the mean HD95, achieving a value of 5.17258 with relatively low standard deviation.

\begin{table}[!h]
    \centering
    \vspace{-5mm}
    \caption{Baselines comparison on the BraTS2018 validation set.}
    \label{tab:brats2018_baselines}
    \begin{tabular}{lcccc}
    \hline
    Methods                     & Mean Dice $\uparrow$ & SD Dice $\downarrow$ & Mean HD95 $\downarrow$ & SD HD95 $\downarrow$ \\ \hline
    CV~\cite{isensee_miccai_2018} & 0.85600 & - & 5.25667 & - \\
    Baseline+BN+DA                        & 0.85568 & 0.00415 & 5.32355 & 0.51249 \\
    \textbf{Baseline}                      & \textbf{0.85394} & \textbf{0.00341} & \textbf{5.70207} & \textbf{0.22982} \\
    Baseline+BN                           & 0.85338 & 0.00320 & 5.17258 & 0.13719 \\
    CV (run once)                        & 0.85089 & - & 5.91061 & - \\
    \hline
    \end{tabular}
    \vspace{-4mm}    
\end{table}

We include the result of Isensee~\emph{et al.} on the BraTS2018 validation set in row four of Table 2~\cite{isensee_miccai_2018}, as it closely aligns with our settings. Their approach demonstrates superior performance due to a 500-epoch training ensemble and 5-fold cross-validation with an ensemble of five models. Adapting the nnU-Net framework's default training scheme~\cite{isensee_nnunet_nature_2021} with a modification of the training epochs to 50 yields a modest mean Dice score of 0.85089 and a mean HD95 of 5.91061. The Baseline+BN+DA variant with intensive data augmentation slightly lags behind Isensee~\emph{et al.}'s approach.

\paragraph{ACS nnU-Net variants:} Table \ref{tab:acs_nnunet} presents the performance of our proposed ACS nnU-Net variants. Both ACS-ResNet18 variants leverage pre-trained weights, achieving comparable performance while significantly reducing the number of parameters. ACS-ResNet18 attains a mean Dice score of 0.85312, while ACS-ResNet18-All outperforms the baseline with a mean Dice score of 0.85444 and a low standard deviation of 0.00130. Pre-trained weights prove vital in leveraging knowledge from natural image datasets, enhancing medical segmentation performance. However, we note that in terms of HD95, we slightly lag behind the baseline.

\begin{table}[!t]
    \centering
    \vspace{-6mm}
    \caption[ACS nnU-Net variants performance]{ACS nnU-Net variants performance on BraTS2018 validation set.}
    \label{tab:acs_nnunet}
    \begin{tabular}{lcccc}
    \hline
    Methods                                & Mean Dice $\uparrow$ & SD Dice $\downarrow$ & Mean HD95 $\downarrow$ & SD Dice $\downarrow$\\ \hline
    ACS-ResNet18-All                        & \textcolor{blue}{\textbf{0.85444}}   & \textcolor{blue}{\textbf{0.00130} } & 5.88153  & 0.34028  \\
    Baseline                                & 0.85394   & 0.00341  & \textcolor{blue}{\textbf{5.70207}}  & \textcolor{blue}{\textbf{0.22982}}  \\
    ACS-ResNet18                            & 0.85312   & 0.00493  & 5.80819  & 0.23196  \\
    \hline
    \end{tabular}
\end{table}

In the competitive context of the BraTS challenge, we evaluate our ACS nnU-Net variants with batch normalization (BN) and intensive data augmentation (DA) in Table \ref{tab:acs_nnunet_bnda}, following a similar approach as Isensee~\emph{et al.}\cite{isensee_nnunet_miccai_2021}. Remarkably, the top-performing model, ACS-ResNet18-All+BN+DA, achieves an impressive mean Dice score of 0.85665 with a low standard deviation of 0.00194, and excels in the mean HD95 with a score of 4.87734. Notably, in both metrics, this variant outperforms the Baseline+BN+DA model and Isensee\emph{et al.}'s ensemble attempt in the BraTS2018 challenge~\cite{isensee_miccai_2018}. The results further validate that BN significantly improves mean HD95 while slightly reducing the mean Dice score, as demonstrated by ACS-ResNet18+BN.

\begin{table}[!t]
    \centering
    \vspace{-6mm}
    \caption{ACS nnU-Net variants with batch normalization and intensive data augmentation performance on BraTS2018 validation set.}
    \label{tab:acs_nnunet_bnda}
    \begin{tabular}{lcccc}
    \hline
    Methods                          & Mean Dice $\uparrow$ & SD Dice $\downarrow$ & Mean HD95 $\downarrow$ & SD Dice $\downarrow$\\ \hline
    ACS-ResNet18-All+BN+DA                  & \textcolor{blue}{\textbf{0.85665}}   & \textcolor{blue}{\textbf{0.00194}}  & \textcolor{blue}{\textbf{4.87734}}  & 0.24654  \\
    CV~\cite{isensee_miccai_2018} & 0.85600  & -   & 5.25667 & - \\
    Baseline+BN+DA                          & 0.85568   & 0.00415  & 5.32355  & 0.51249  \\
    ACS-ResNet18+BN+DA                      & 0.85389   & 0.00324  & 5.42373  & 0.65147  \\
    Baseline+BN                             & 0.85338   & 0.00320  & 5.17258  & \textcolor{blue}{\textbf{0.13719}}  \\
    ACS-ResNet18+BN                         & 0.84999   & 0.00233  & 5.22337  & 0.13054  \\
    \hline
    \end{tabular}
    \vspace{-4mm}
\end{table}

\vspace{-2mm}
\paragraph{JCS nnU-Net:}
Table \ref{tab:jcs} presents a detailed performance comparison of the proposed JCS nnU-Net and HGG-LGG nnU-Net against the baseline. The HGG-LGG nnU-Net achieves an impressive mean Dice score of 0.8071 in the enhancing tumor region, aligning with its classification proxy task of distinguishing high-grade gliomas (HGG) from low-grade gliomas (LGG). However, there is an unexpected degradation in the tumor core region, indicating limitations in capturing complex features specific to TC segmentation, resulting in lower overall performance compared to the Baseline in terms of mean Dice score.

In contrast, the JCS nnU-Net improves over the baseline, achieving a mean Dice score of 0.8557 compared to the baseline's 0.8539. Notably, the JCS nnU-Net outperforms the HGG-LGG nnU-Net in most sub-regions, except for the enhancing tumor (ET) region where the HGG-LGG nnU-Net excels, leveraging the strengths of both classification and segmentation tasks. The JCS approach proves to be a powerful and balanced strategy, effectively combining pre-existing knowledge from classification with fine-tuned segmentation to achieve superior performance.

\begin{table}[!t]
    \centering
    \vspace{-4mm}
    \caption{Performance of JCS nnU-Net and HGG-LGG nnU-Net. Abbreviations: whole tumor (WT), tumor core (TC), and enhancing tumor (ET). Four floating point precision is used for clarity to avoid crowding the table with numbers.}
    \label{tab:jcs}
    \begin{tabular}{l|cccc|cccc}
    \hline
     & \multicolumn{4}{c}{Mean Dice $\uparrow$} & \multicolumn{4}{|c}{Mean HD95 $\downarrow$} \\ \hline
    Methods  & WT      & TC     & ET & All      & WT      & TC     & ET & All\\ \hline
    JCS      & \textcolor{blue}{\textbf{0.9110}} & \textcolor{blue}{\textbf{0.8546}} & 0.8016 & \textcolor{blue}{\textbf{0.8557}} & 4.7712 & \textcolor{blue}{\textbf{7.055}} & 5.2610 & \textcolor{blue}{\textbf{5.6958}} \\ 
    Baseline & 0.9105 & 0.8508 & 0.8005 & 0.8539 & 4.5199  & 7.4927 & \textcolor{blue}{\textbf{5.0937}} & 5.7021 \\  
    HGG-LGG nnU-Net  & \textcolor{blue}{\textbf{0.9110}} & 0.8429 & \textcolor{blue}{\textbf{0.8071}} & 0.8537 & \textcolor{blue}{\textbf{4.4473}} & 8.1035 & 5.1186 & 5.8898 \\
    \hline
    \end{tabular}
\end{table}

\vspace{-2mm}
\paragraph{Post-processing results:}
We evaluate our proposed methods with post-processing techniques, as shown in Table \ref{tab:2018_result_post}. The ACS-ResNet18+BN+DA+Post and ACS-ResNet18-All+BN+DA+Post variants achieve the highest mean Dice scores of 0.86231 and 0.86170, respectively, along with impressive segmentation performance in terms of mean HD95. Interestingly, the ACS-ResNet18+BN+Post variant outperforms Baseline+BN+DA+Post, indicating that using BN and post-processing alone already achieves competitive performance, considering the significant increase in training time associated with DA. Notably, all post-processed methods surpass the default nnU-Net training baseline with an ensemble of 5 models in cross-validation, as well as Isensee~\emph{et al.}'s previous attempt~\cite{isensee_miccai_2018}, in terms of Mean Dice score.

\begin{table}[!t]
    \centering
    \vspace{-4mm}
    \caption{Results on BraTS2018 validation set.}
    \label{tab:2018_result_post}
    \begin{tabular}{l|cc|cc}
    \hline
    Methods                  & Mean Dice $\uparrow$ & SD Dice $\downarrow$ & Mean HD95 $\downarrow$ & SD HD95 $\downarrow$\\ \hline
    ACS-ResNet18+BN+DA+Post      & \textcolor{blue}{\textbf{0.86231}} & 0.00062 & 4.95997 & 0.20410 \\
    ACS-ResNet18-All+BN+DA+Post  & 0.86170   & 0.00194  & \textcolor{blue}{\textbf{4.87734}}  & 0.24654 \\
    JCS+Post                  & 0.86098   & 0.00203  & 5.43545 & 0.23385 \\
    ACS-ResNet18+BN+Post         & 0.86009   & 0.00233  & 4.88076 & \textcolor{blue}{\textbf{0.13614}} \\
    Baseline+Post             & 0.85993   & \textcolor{blue}{\textbf{0.00045}}  & 5.47518 & 0.20718 \\
    Baseline+BN+DA+Post       & 0.85899   & 0.00163  & 4.93538 & 0.28096 \\
    CV~\cite{isensee_miccai_2018} & 0.85600  & -      & 5.25667 & - \\
    Baseline+CV (run once)     & 0.85089   & -      & 5.91061 & - \\
    \hline
    \end{tabular}
\end{table}

\vspace{-2mm}
\subsubsection{BraTS2020:}
\begin{table}[!b]
    \centering
    \vspace{-4mm}
    \caption{Results on BraTS2020 validation set.}
    \label{tab:2020_result_post}
    \begin{tabular}{lcc}
    \hline
    Methods                 & Mean Dice $\uparrow$ & Mean HD95 $\downarrow$\\ \hline
    CV~\cite{isensee_nnunet_miccai_2021} & \textcolor{blue}{\textbf{0.85580}} & 11.93000 \\
    ACS-ResNet18+BN+DA+Post & 0.85546           & \textcolor{blue}{\textbf{11.44526}} \\
    JCS+Post                & 0.84291           & 17.05532 \\
    JCS                     & 0.83009           & 17.31020 \\
    ACS-ResNet18+BN+DA      & 0.82553           & 19.71874 \\
    \hline
    \end{tabular}
\end{table}

We validate the relevance and adaptability of our methods on the BraTS2020 dataset~\cite{menze_tmi_2015,bakas_arxiv_2019} and compare them with the winning solution by Isensee~\emph{et al.}~\cite{isensee_nnunet_miccai_2021} in BraTS2020, which shares similar training settings. Table \ref{tab:2020_result_post} presents the results, highlighting the performance of our approaches.

Our ACS-ResNet18+BN+DA+Post model achieves a mean Dice score of 0.85546, closely approaching the best-performing CV model with a Dice score of 0.85580. Additionally, our model performs competitively in terms of the mean 95\% Hausdorff distance, with a value of 11.44526, compared to the winning solution's 11.93. The JCS nnU-Net model does not show significant improvement with post-processing, with a mean Dice score of 0.84291 and a marginal increase in the mean HD95 (17.31020). In contrast, after applying post-processing, our ACS nnU-Net variants demonstrate substantial improvements in both metrics.

\subsection{Qualitative results}
\label{sec:qual_results}

\subsubsection{BraTS2018}
\label{subsec:qual_brats2018}

\begin{figure}[!ht]
    \centering
    \begin{subfigure}{0.65\textwidth}
        \centering
        \includegraphics[width=\textwidth]{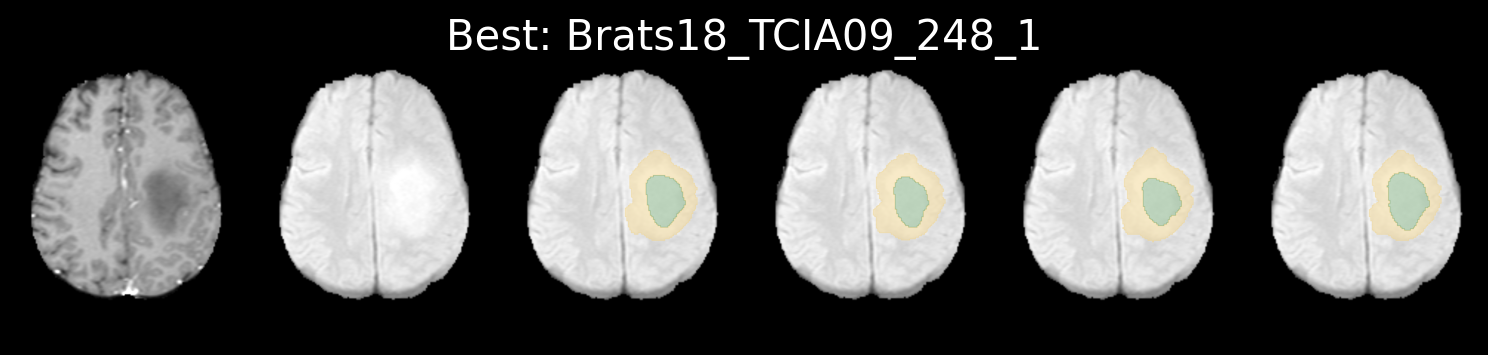}
    \end{subfigure}
    \begin{subfigure}{0.65\textwidth}
        \centering
        \includegraphics[width=\textwidth]{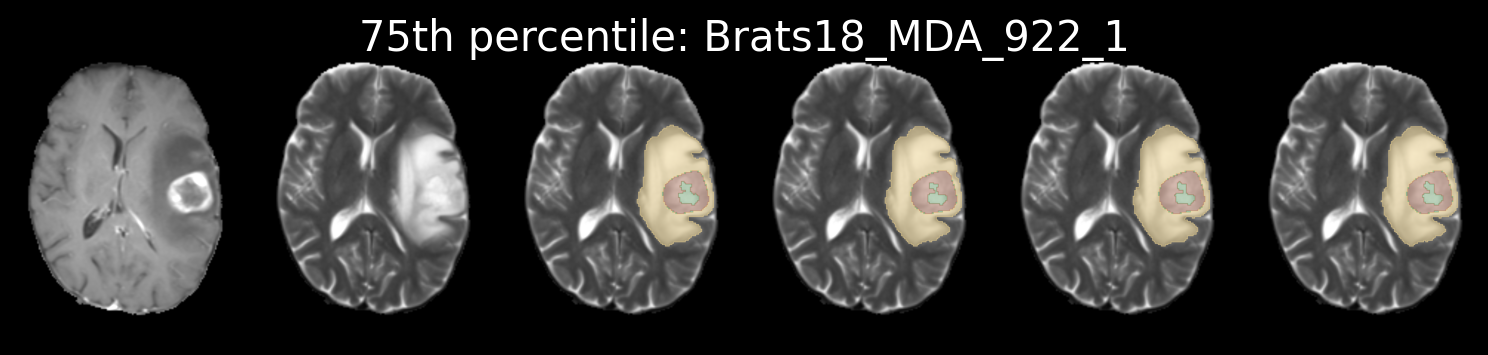}
    \end{subfigure}
    \begin{subfigure}{0.65\textwidth}
        \centering
        \includegraphics[width=\textwidth]{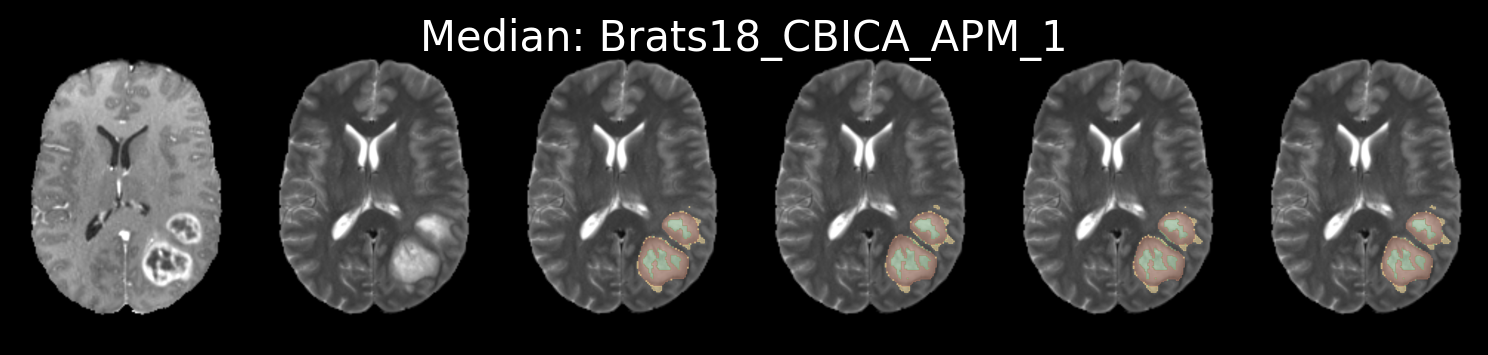}
    \end{subfigure}
    \begin{subfigure}{0.65\textwidth}
        \centering
        \includegraphics[width=\textwidth]{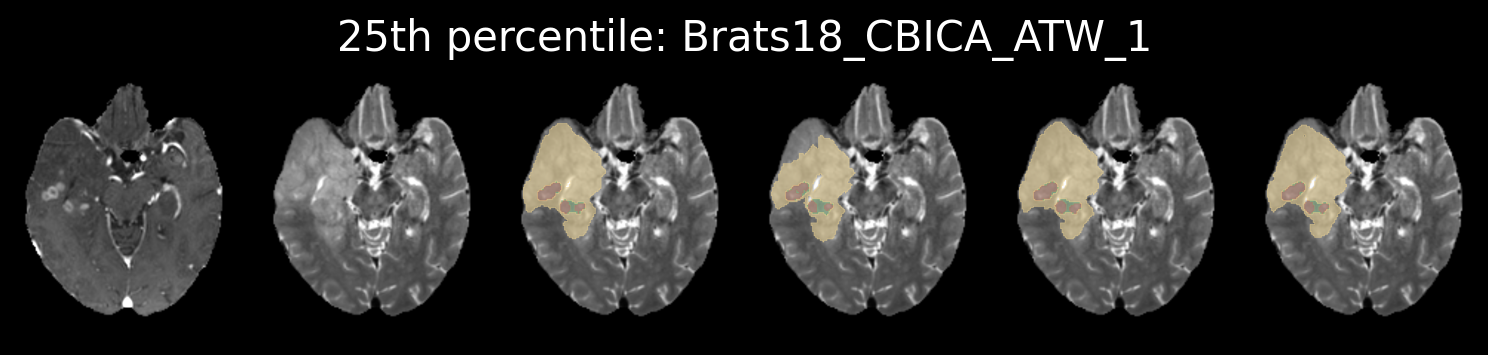}
    \end{subfigure}
    \begin{subfigure}{0.65\textwidth}
        \centering
        \includegraphics[width=\textwidth]{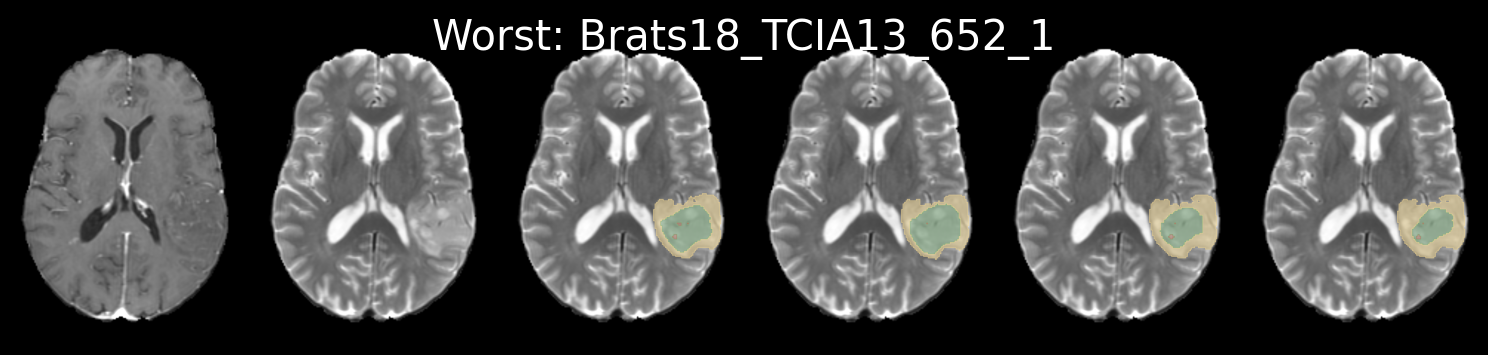}
    \end{subfigure}
    \caption[Qualitative BraTS2018 validation set results]{Qualitative results on the BraTS2018 validation set: Rows show the best, 75th percentile, median, 25th percentile, and worst predictions of our best model ACS-ResNet18+BN+DA+Post. Columns display raw T1ce, raw T2, Baseline+BN+DA+Post, JCS+Post, ACS-ResNet18-All+BN+DA+Post, and ACS-ResNet18+BN+DA+Post predictions. Yellow, green, and red highlight the whole tumor, tumor core, and enhancing tumor, respectively.}
    \label{fig:val_visualize_2018}
\end{figure}

Figure \ref{fig:val_visualize_2018} provides a qualitative overview of the segmentation performance of our top 3 models, along with the baseline, after pre-processing. The first row demonstrates the effectiveness of post-processing in removing insignificant enhancing tumor voxels, resulting in the display of only whole tumor (WT) and necrosis (NCR) labels. As ground truth masks are unavailable for direct comparison, we highlight that the ACS-based nnU-Net variants, despite having fewer parameters, exhibit competitive segmentation performance compared to the baseline nnU-Net.

\subsubsection{BraTS2020:}
\begin{figure}[t]
    \centering
    \begin{subfigure}{0.42\textwidth}
        \centering
        \includegraphics[width=\textwidth]{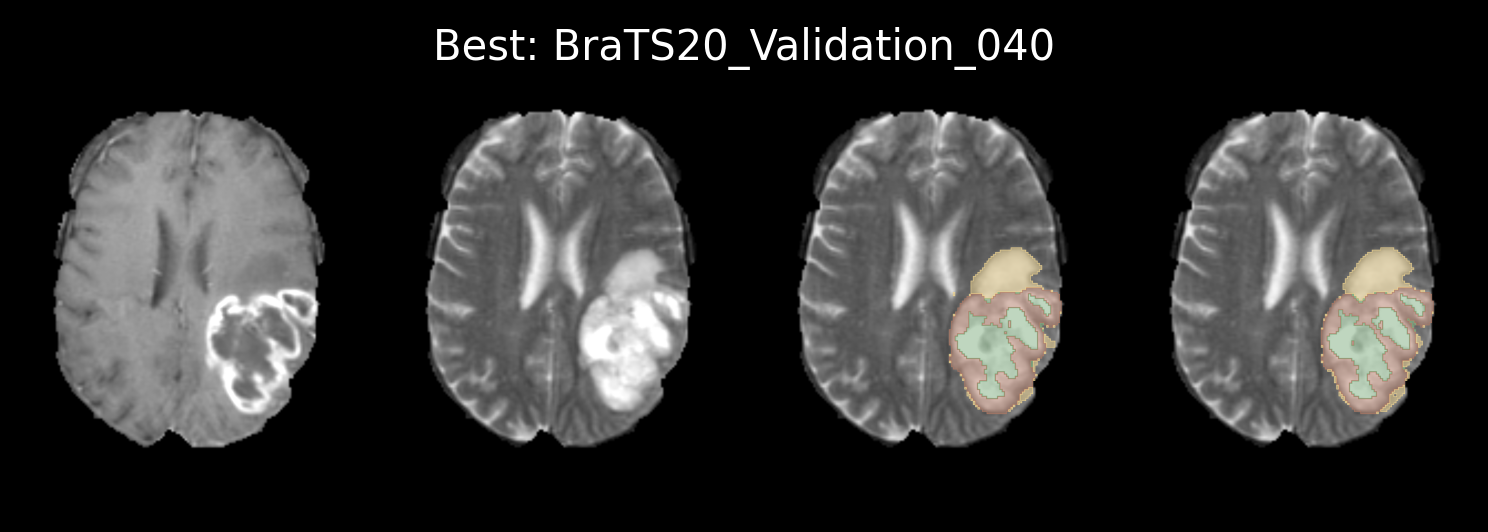}
    \end{subfigure}
    \begin{subfigure}{0.42\textwidth}
        \centering
        \includegraphics[width=\textwidth]{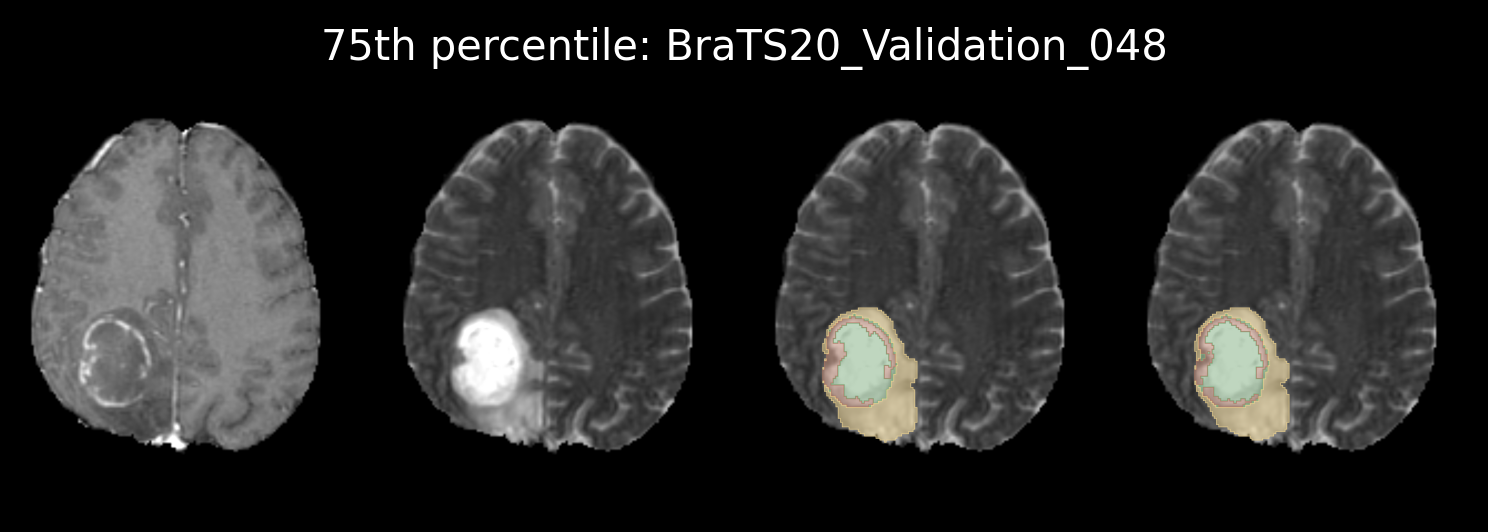}
    \end{subfigure}
    \begin{subfigure}{0.42\textwidth}
        \centering
        \includegraphics[width=\textwidth]{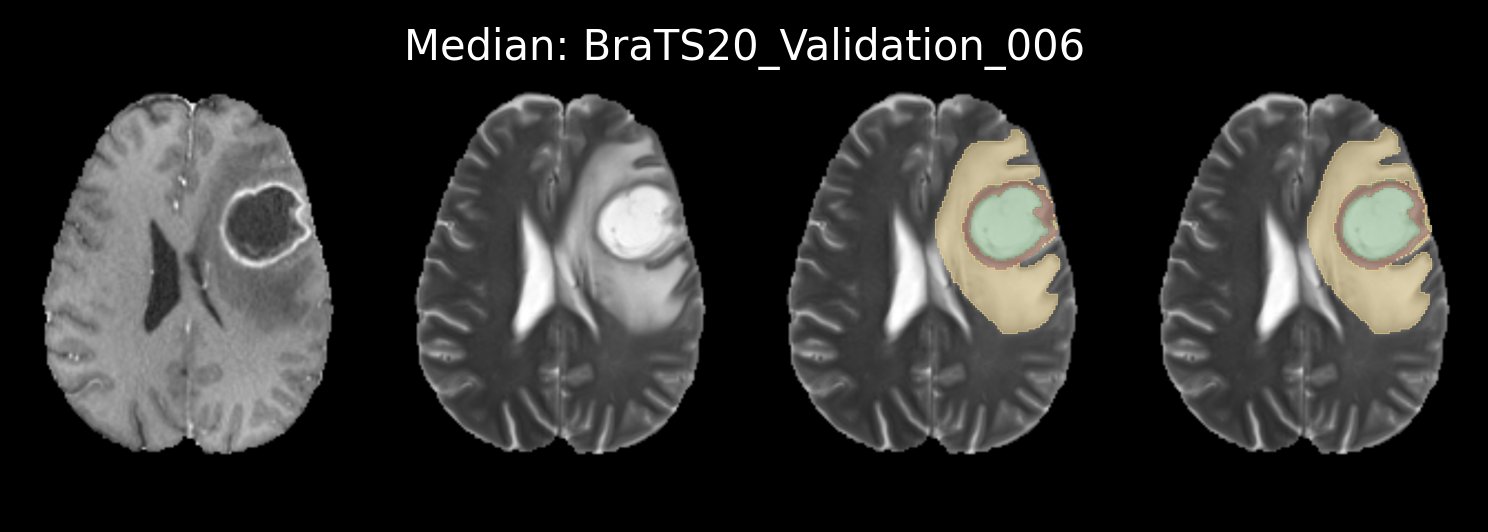}
    \end{subfigure}
    \begin{subfigure}{0.42\textwidth}
        \centering
        \includegraphics[width=\textwidth]{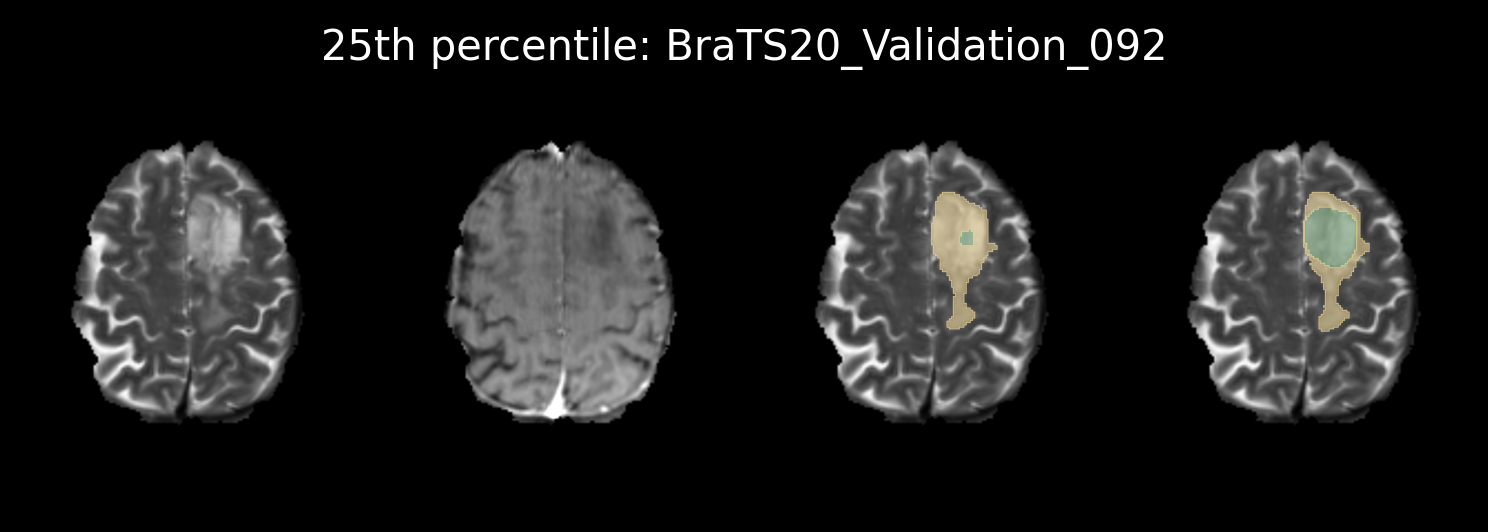}
    \end{subfigure}
    \begin{subfigure}{0.42\textwidth}
        \centering
        \includegraphics[width=\textwidth]{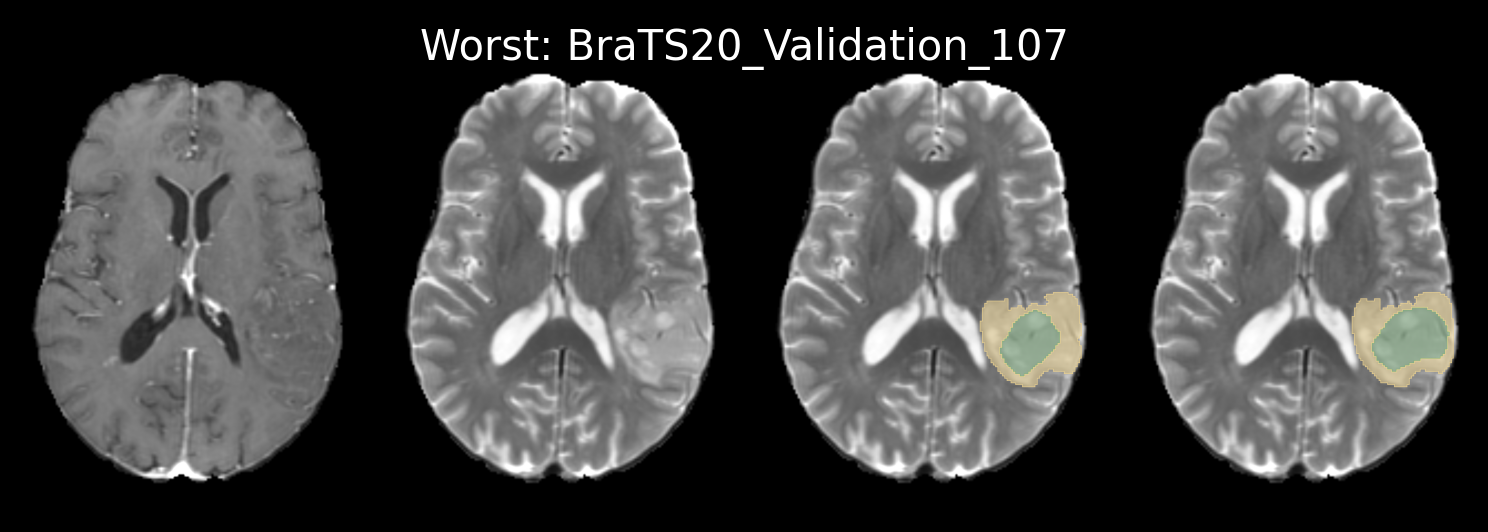}
    \end{subfigure}
    \caption{Qualitative results on the BraTS2020 validation set: Rows show the best, 75th percentile, median, 25th percentile, and worst predictions of our best model ACS-ResNet18+BN+DA+Post. Columns display raw T1ce, raw T2, JCS+Post, and ACS-ResNet18+BN+DA+Post predictions. Yellow, green, and red highlight the whole tumor, tumor core, and enhancing tumor, respectively.}
    \label{fig:val_visualize_2020}
    \vspace{-4mm}
\end{figure}

Figure \ref{fig:val_visualize_2020} presents qualitative results of our proposed methods on the updated BraTS2020 dataset. The 25th percentile case indicates the successful removal of negligible ET voxels through post-processing. Additionally, the ACS-based approach outperforms JCS in the TC region significantly.

\subsection{Space and time complexity analysis}
The baseline nnU-Net, with 88.629 million parameters, achieves a relatively quick inference time of 4.7776 seconds. On the other hand, the JCS nnU-Net model, incorporating additional adapters in each stage, has 125.158 million parameters and requires 7.4381 seconds for inference. The ACS-based nnU-Net stands out with only 18.601 million parameters, which is approximately 21\% of the baseline's parameter count; however, the inference time is 6.4320 seconds. While ACS convolutions offer efficiency benefits, their current PyTorch implementation lacks native support, indicating a need for more efficient implementations to achieve competitive results in terms of runtime~\cite{yang_acs_jbhi_2021}.

\begin{table}[!h]
    \vspace{-7mm}
    \centering
    \caption[Trainable parameters and inference time comparison]{Trainable parameters and inference time comparison.}
    \begin{tabular}{lcc}
    \hline
    Methods           & Trainable parameters $\downarrow$ & Inference time (s) $\downarrow$ \\ \hline
    ACS nnU-Net & \textcolor{blue}{\textbf{18.601M}}       & 6.4320 \\
    Baseline    & 88.629M       & \textcolor{blue}{\textbf{4.7776}} \\
    JCS nnU-Net & 125.158M      & 7.4381 \\
    \hline
    \end{tabular}
    \vspace{-4mm}    
    \label{tab:params_flops}
\end{table}

%% file: mainchaps/conclusion.tex
\vspace{-4mm}
\section{Conclusion}
\vspace{-1mm}
In this paper, we present two efficient brain tumor segmentation approaches within the fast training settings of the nnU-Net framework. By seamlessly integrating 2D pre-trained weights using ACS convolutions into the 3D domain, we reduce trainable parameters, enhance resource efficiency, and maintain robust segmentation performance. The Joint Classification and Segmentation (JCS) model shows promise, despite its current underperformance, as it combines classification and segmentation tasks. Our methods are evaluated on the BraTS2018 and BraTS2020 datasets. The extensive experiments show the competitive performance of our proposed work. 

Future work involves exploring the generalization of ACS convolutions' pre-trained weights to nnU-Net architectures and optimizing the JCS approach. We also aim to evaluate our proposed method on the later BraTS datasets in the future.

\textbf{Acknowledgment:} This research is funded by Viet Nam National University Ho Chi Minh City (VNU-HCM) under grant number DS2020-42-01.